\documentclass[fleqn,12pt,twoside]{article}
\usepackage[headings]{espcrc1}
\usepackage{wrapfig}
\usepackage{graphicx}
\title{First Measurement of the $\omega$-meson Production at RHIC by PHENIX}

\author{V.Ryabov (for the PHENIX\footnote{For the full list of PHENIX authors and aknowledgements, see appendix 'Collaborations' of this volume.}
Collaboration)\address[MCSD]{Peterburg Nuclear Physics Institute, Gatchina, 188300, Russian Federation\\
e-mail:ryabovvg@mail.pnpi.spb.ru}}

\runtitle{First Measurement of the $\omega$-meson Production at RHIC by PHENIX}
\runauthor{V.Ryabov (for the PHENIX Collaboration)}

\begin{document}

\maketitle

\begin{abstract}
The PHENIX~\cite{phenix} experiment at RHIC measured the $\omega$-meson production in the range of $p_{\perp}$ from 2.5~$GeV/c$ to 9.5~$GeV/c$ in $p+p$ and $d+Au$ collision systems at $\sqrt{s_{NN}}=200$~$GeV$. We present the consistent results of two independent measurements in decay channels $\omega\rightarrow\pi^{0}\pi^{+}\pi^{-}$ and $\omega\rightarrow\pi^{0}\gamma$. The $\omega/\pi^{0}$ ratio was found to be $0.90\pm0.06$. The Nuclear Modification Factor $R_{dA}$ is $1.3\pm 0.2$ both in the Minimum Biased and (0\%-20\%) central event samples. The mass of the $\omega$-meson was measured to be consistent with the PDG\cite{PDG} value within the errors of the measurement. We also measured the $K_{S}^{0}$-meson production in $K_{S}^{0}\rightarrow\pi^{0}\pi^{0}$ in the same collision systems.
\end{abstract}

\section{INTRODUCTION}
The $\omega$-meson carries vast information about the physics of the Heavy Ion Collisions. Ratio of vector to pseudo-scalar meson yields such as $\omega/\pi^{0}$ is informative about the production of $q\bar{q}$ bound states. Nuclear Modification Factor $R^{\omega}_{dA}$ completes the picture of other meson $R_{dA}$ at the mass point of ($m_{\omega}=782$~$MeV$). $\omega$-meson is among the most sensitive probes of the media properties. Due to its large width ($\Gamma=8.5$~$MeV$) a significant fraction of $\omega$-mesons decay inside the dense (or hot and dense) matter present in the collision. Mesons produced in collsions of heavy nuclei can have different mass and width compared to the mesons created in vacuum. Comparison of $\omega$-meson yields in different channels in different collisons systems is another way to address the same question. Current results can be used as a baseline for such studies.


\section{ANALYSIS}
The $\omega$-meson is abundantly produced in the HI collisions, but so far its measurement was precluded by combinatorial background coming from the $\pi$-mesons. In this analysis we took the advantage of the excellent PHENIX trigger capability to compensate for rather limited geometrical acceptance of the detector. Triggering on energetic photons enabled us for the first time in PHENIX to study the dominant three body decay mode $\omega\rightarrow\pi^{0}\pi^{+}\pi^{-}\rightarrow 2\gamma 2\pi$. Complemented with the two body decay mode $\omega\rightarrow\pi^{0}\gamma \rightarrow 3\gamma$ results and additional mode of $K_{S}^{0}\rightarrow\pi^{0}\pi^{0}\rightarrow 4\gamma$ we obtained sets of measurements which can be verified for consistency by comparision with each other and with previously published data.

The analysis is based on data samples of $4.6\times10^{7}$($2.1\times10^{7}$)\footnote{Here and below the first number corresponds to $p+p$ sample, number in brackets to $d+Au$} triggered events passed which Quality Assurance control. Each event satisfies the Minimum Biased ($MB$) requirement of at least one hit in North and South Beam-Beam Counters. We selected events with vertex within 30~$cm$ of the detector center and at least one energetic photon in either of the PHENIX Central Arms. The latter requirement imposed online allowed to record the event sample equavalent to $5\times10^{9}$($3\times10^{9}$) MB collisions.

\begin{figure}[htb]
\begin{center}
\vspace{-5mm}
\includegraphics[width=0.4\linewidth]{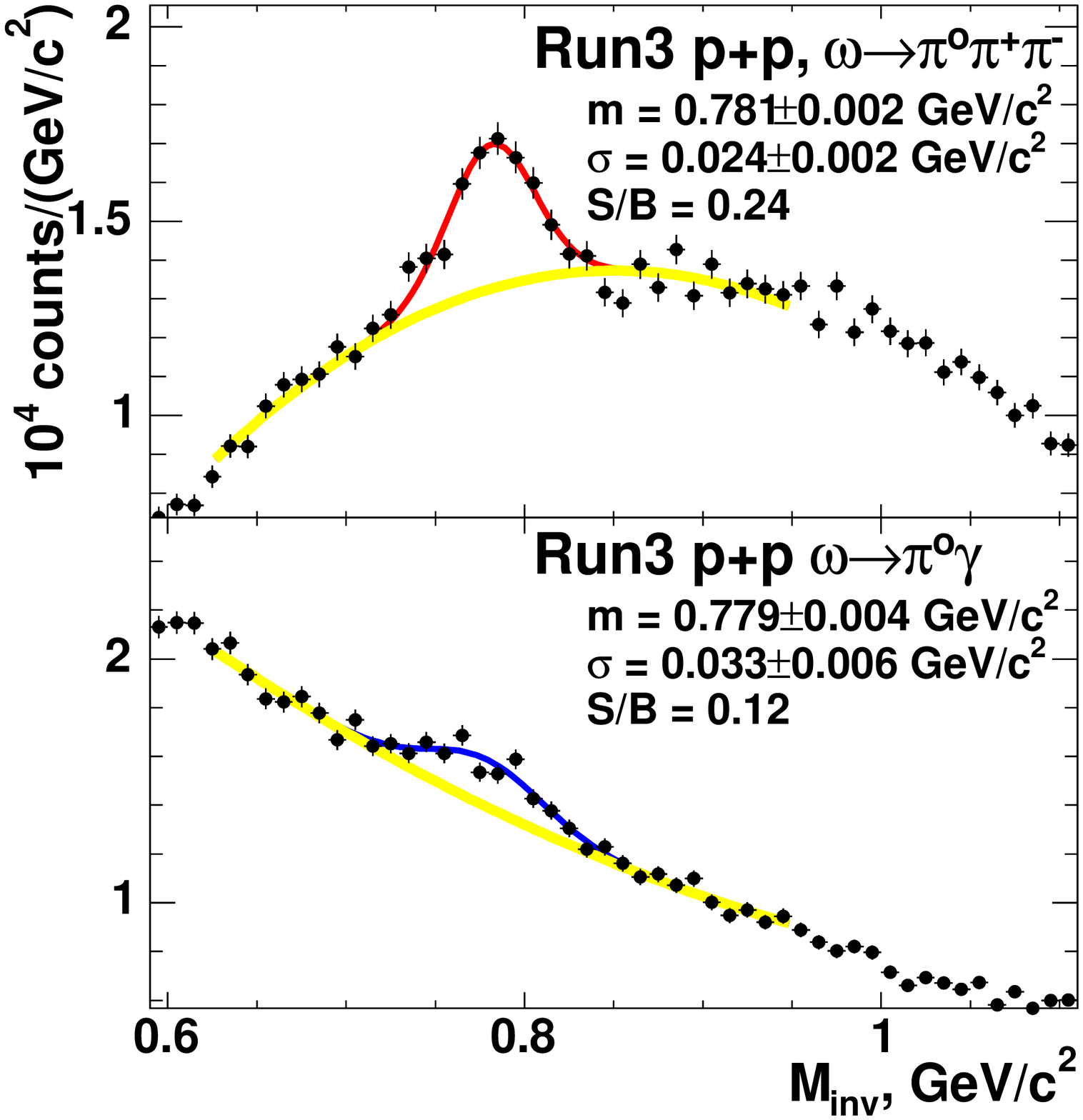}
\includegraphics[width=0.4\linewidth]{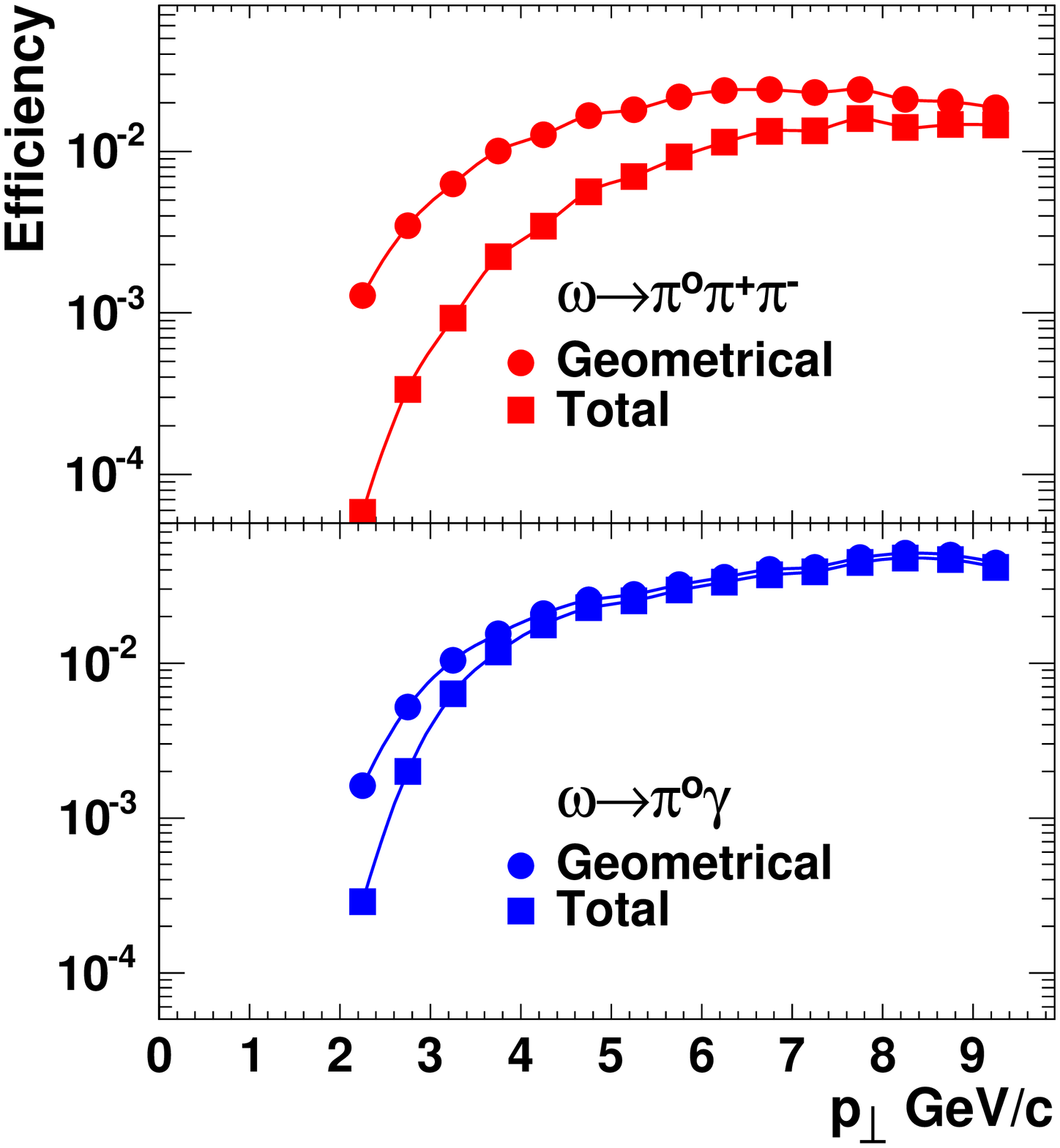}
\end{center}
\vspace{-12mm}
\caption{Left: $\omega$-meson peaks in the invariant mass distribution of  $\pi^{0}\pi^{+}\pi^{-}$ and $\pi^{0}\gamma$. Right: PHENIX geometrical acceptance and total acceptance including trigger efficiency.\label{fig:peaks}}
\vspace{-7mm}
\end{figure}

First the $\pi^{0}$-mesons candidates are reconstructed using PHENIX calorimeter. Particles with reconstructed mass within 2 standard deviations ($p_{\perp}$ dependent) are accepted and combined depending on the channel of study with a single $\gamma$ (BR 8.5\%), $\pi^{+}\pi^{-}$ pair (BR 89\%) or another $\pi^{0}$ candidate for $K_{S}^{0}$ (BR 31\%). Reconstructed peaks for $\omega$-meson decays\footnote{Online version of this procceding shows $\omega\rightarrow\pi^{0}\pi^{+}\pi^{-}$ results in red and $\omega\rightarrow\pi^{0}\gamma$ in blue} are shown in the left panel of Figure~\ref{fig:peaks}. 
Integrals under peaks are extracted by fitting. Mixed event beckground subtraction fails due to particle correlations present in the direct event. A 10\% systematic error accociated with yield extraction was deduced by varying the analysis cuts and fitting options. Raw yields are corrected for trigger efficiencies and geometrical acceptance. The corrections are shown in the right panel of Figure~\ref{fig:peaks}. Acceptance and trigger corrections are obtained by simulation based on full detector description, on-line trigger settings of 1.4~$GeV$(2.4~$GeV$) in the calorimeter, and decay kinematics of a particular decay channel. For the decay channel $\omega\rightarrow\pi^{0}\pi^{+}\pi^{-}$ the latter takes into account phase-space density distribution published in~\cite{dalitz1,dalitz2}.

We consider systematic errors related to reconstruction of neutral and charged particles, errors related to trigger efficiency determination and analysis uncertainties. Different sources dominate the systematic errors at different $p_{\perp}$. Major contributions are coming from the MB trigger accuracy, calorimeter energy scale and extraction of raw yields. Total systematic error for decays under study is in the range 15\%-25\%. PHENIX capability to measure different particles species within the same data sample allows to reduce systematic errors in ratios such as $\omega/\pi^{0}$, $R_{dA}$.

\section{RESULTS}
Invariant yields of $\omega$-meson in $p+p$, MB and (0\%-20\%) $d+Au$ at $\sqrt{s_{NN}}=200$~$GeV$ 
\begin{wrapfigure}{l}{9cm}
\vspace{-7mm}
\includegraphics[width=1.0\linewidth]{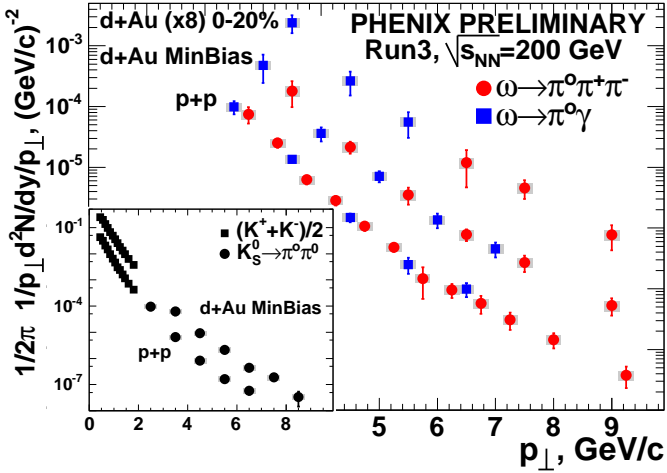}
\vspace{-5mm}
\caption{Invariant yields of $\omega$-meson at $\sqrt{s_{NN}}=200$~$GeV$. Insert shows $K_{S}^{0}\rightarrow\pi^{0}\pi^{0}$ results.\label{fig:spectra}}
\vspace{-8mm}
\end{wrapfigure}
measured in two decay channels are shown in Figure~\ref{fig:spectra}. Result for $\omega\rightarrow\pi^{0}\pi^{+}\pi^{-}$ and $\omega\rightarrow\pi^{0}\gamma$ are consistent with each other. The $p_{\perp}$-range extends from 2.5~$GeV/c$ to 9.5~$GeV/c$ and is limited by statistics on higher side and by photon trigger efficency on lower side. Lower $p_{\perp}$ can be achieved by using MB event sample. In this case the geometrical acceptance becomes the next limiting factor as can be seen from Figure~\ref{fig:peaks}. The $K_{S}^{0}$ results presented in the insert cover $p_{\perp}$ (2.5-8.5)~$GeV/c$ and complement previous PHENIX measurement of $K^{+},K^{-}$~\cite{felix} done at lower $p_{\perp}$. The data are consistent with STAR results of $K_{S}^{0}\rightarrow\pi^{+}\pi^{-}$~\cite{starK} in the region of overlap and extends further in $p_{\perp}$. 

One of the most interesting results is the ratio of the vector to pseudo-scalar meson \begin{wrapfigure}{l}{9cm}
\vspace{-7mm}
\includegraphics[width=1.\linewidth]{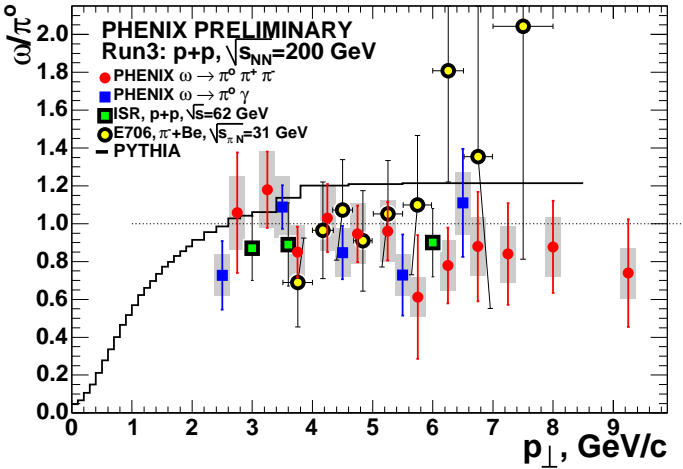}
\vspace{-5mm}
\caption{$\omega/\pi^{0}$ vs. $p_{\perp}$ compared to IR and E706 measurements and PYTHIA predictions.\label{fig:ratio}}
\vspace{-8mm}
\end{wrapfigure} yields at high $p_{\perp}$. Figure~\ref{fig:ratio} presents the first result on that observable measured at $\sqrt{s}=200$~$GeV$ in $p+p$ collisions together with the lower energy measurements. At RHIC $\omega/\pi^{0}$ ratios reaches $\sim0.9$ above $p_{\perp}>3.5$~$GeV/c$. PYTHIA~\cite{pythia} prediction at the same energy is higher, but consistent with it within errors. One should note that PHENIX preliminary result is not corrected for feed-down into $\pi^{0}$ yield. At $\sqrt{s_{\pi N}}=31$~$GeV$ the E706~\cite{e706} experiment measured the same ratio to be growing with $p_{\perp}$, while at $\sqrt{s}=62$~$GeV$ the ISR~\cite{isr} collaboration found it flat at the same magnitude.

\begin{figure}[htb]
\vspace{-5mm}
\begin{minipage}[t]{92mm}
\includegraphics[width=1.0\linewidth]{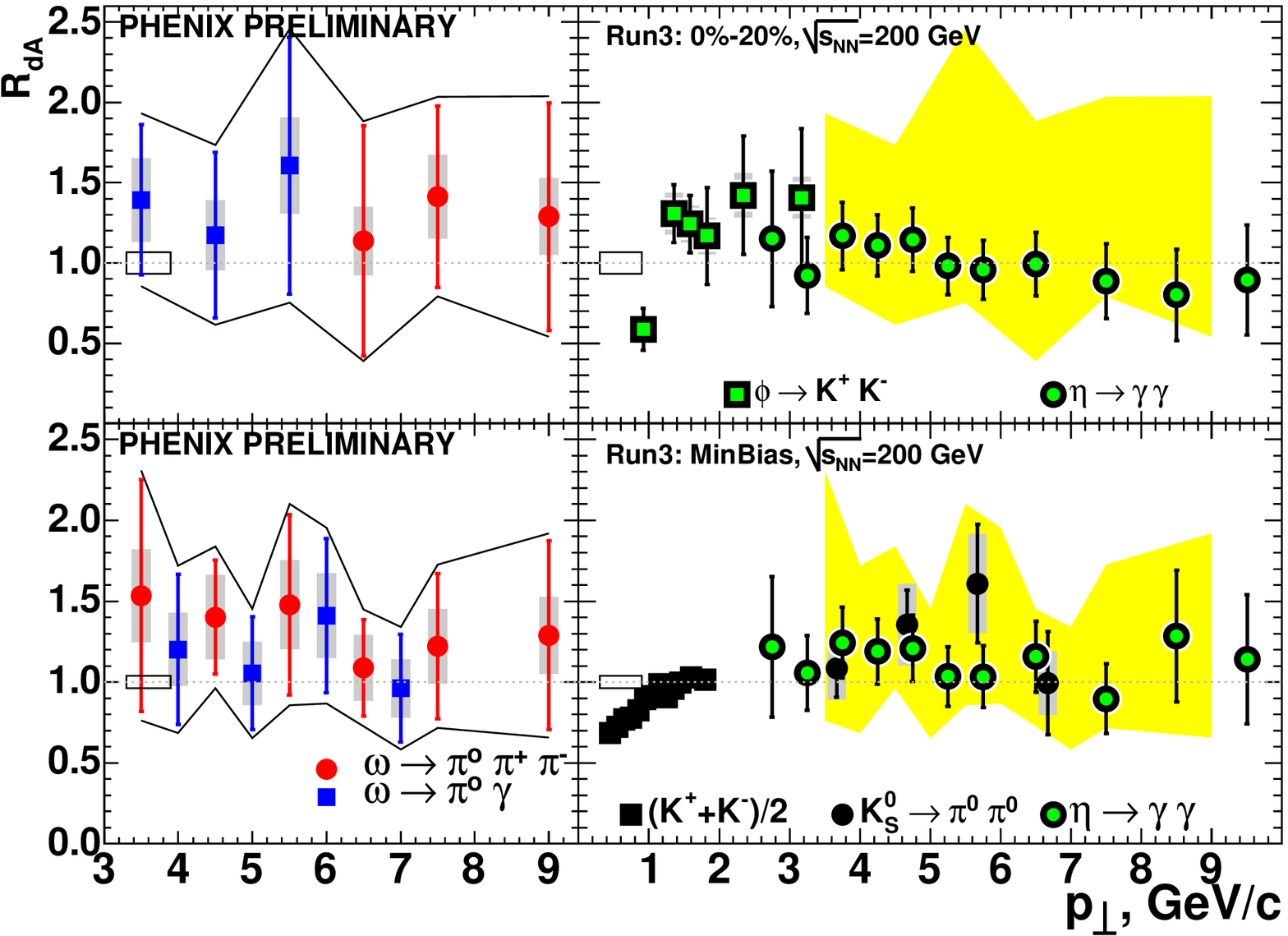}
\vspace{-12mm}
\caption{$R^{\omega}_{dA}$ vs. $p_{\perp}$ compared to the results other neutral mesons: $R^{\eta}_{dA}$, $R^{K}_{dA}$, and $R^{\phi}_{CP}$.\label{fig:rda}}
\end{minipage}
\hspace{\fill}
\begin{minipage}[t]{65mm}
\includegraphics[width=1.0\linewidth]{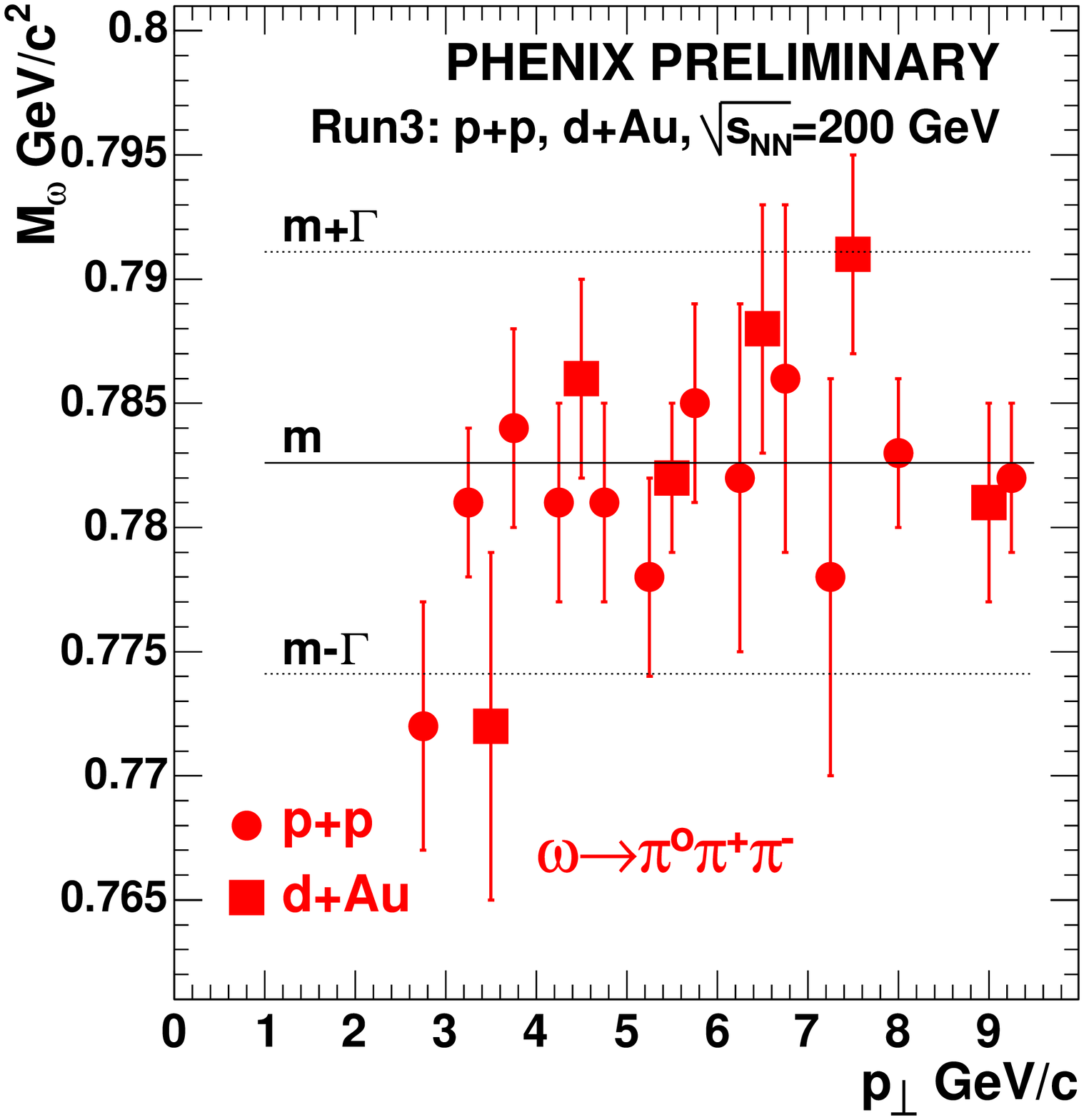}
\vspace{-12mm}
\caption{Reconstructed mass of the $\omega\rightarrow\pi^{0}\pi^{+}\pi^{-}$ vs. $p_{\perp}$.\label{fig:mass}}
\end{minipage}
\vspace{-5mm}
\end{figure}

Results on $\omega$-meson $R_{dA}$ are shown in Figure~\ref{fig:rda} (left). We observe $R^{\omega}_{dA}$ to be flat over $p_{\perp}$-range of the measurement both in the MB and (0\%-20\%) central event samples and equal to $1.3 \pm 0.2$. This number is consistent within errors with previously reported preliminary results of PHENIX on $R_{dA}$ of $\eta$-mesons~\cite{eta-pi} shown in Figure~\ref{fig:rda} (right) on top of the band of $R^{\omega}_{dA}$ results from the left panel. $R^{\omega}_{dA}$ is very close to $R_{dA}$ of $K$-meson measured in the framework of this analysis and $R_{CP}$ of $\phi$-meson~\cite{phi}.

There is special interest to the properties of reconstructed $\omega$-meson due to their possible modification inside the media because a significant fraction of mesons decay very early. This can affect the mass peak position and the width of the peaks shown in Figure~\ref{fig:peaks}. Some recent publications suggest that the $\omega$-meson mass modification can be observed in cold matter by studying not only electron decay channel~\cite{KEK} but hadronic channels~\cite{mass}. PHENIX, lacking acceptance at low $p_{\perp}$ where the effect is the most prominent nevertheless has very fine mass resolution. Figure~\ref{fig:mass} demonstrates that even with rather limited statistics one can measure $p_{\perp}$-dependence of the $\omega$-meson mass. In the range of the measurement available in this analysis it is found to be the same in $p+p$ and $d+Au$ collision and consistent with the PDG value. Possibility to measure the width of the meson is under investigation. This result would also allow a direct comparison with the $\omega\rightarrow e^{+}e^{-}$ channel in different collision systems to be done in future.

\end{document}